\documentclass{mem}
\usepackage{natbib}
\usepackage{txfonts}
\usepackage{balance}
\usepackage{graphicx}
\usepackage[hidelinks]{hyperref}
\usepackage{amsmath}
\usepackage{amssymb}	
\usepackage[T1]{fontenc}
\usepackage{ae,aecompl}
\usepackage{multirow}
\usepackage{tabularx}
\bibpunct{(}{)}{;}{a}{}{,} 
\usepackage{siunitx}
\newcommand{\xmm}{\textit{XMM-Newton}}
\newcommand{\nustar}{\textit{NuSTAR}}
\newcommand{\rxte}{\textit{Rossi-XTE}}
\newcommand{\swift}{\textit{Swift}}

\newcommand{\beppo}{\textit{BeppoSAX}}
\newcommand{\source}{\rm{MXB 1659-298}}
\begin{document}

\title{Broadband spectral analysis of MXB 1659-298 in its soft and hard state}


\author{S. M. Mazzola\inst{1}\thanks{\email{simonamichela.mazzola@unipa.it}}, R. Iaria\inst{1},  A. F. Gambino\inst{2}, A. Marino\inst{1,3,4},  T. Di Salvo\inst{1},  T. Bassi\inst{1,3,4},  A. Sanna\inst{2},  A. Riggio\inst{2}, L. Burderi\inst{2}}

\institute{Dipartimento di Fisica e Chimica - Emilio Segr\`e, Universit\`a degli Studi di Palermo, via Archirafi 36 - 90123 Palermo, Italy
\and
Universit\`a
  degli Studi di Cagliari, Dipartimento di Fisica, SP
  Monserrato-Sestu, KM 0.7, 09042 Monserrato, Italy\label{inst2}\and 
INAF -- Istituto di Astrofisica Spaziale e Fisica Cosmica di Palermo,
Via Ugo La Malfa 153, I-90146 Palermo, Italy\label{inst3}\and  
IRAP, Universit\'{e} de Toulouse, CNRS, UPS, CNES, Toulouse, France\label{inst4} }

\authorrunning{S. M. Mazzola et al.}
\titlerunning{Broadband spectral analysis of MXB 1659-298}
\abstract{
The X-ray transient eclipsing source \source \ went in outburst in 1999 and 2015, respectively, during which it was observed by \xmm, \nustar \ and \swift.
Using these observations we studied the broadband spectrum of the source to constrain the continuum components and to verify the presence of a reflection component.
We analysed the soft and hard state of the source, finding that the soft state can be modelled with a thermal component associated with the inner accretion disc plus a Comptonised component. A smeared reflection component and the presence of an ionised absorber are also requested in the best-fit model.
On the other hand, the direct continuum emission in the hard state can be described by a Comptonised component with a temperature larger than 150 keV. Also in this case a reflection component and a ionised absorber are observed.
}
\maketitle{}

\section{Introduction}
\source\ is a low mass X-ray binary (LMXB) harbouring a neutron star (NS). 
The source was observed in outburst three times: from 1976 up to 1978 \citep{Com_83,Com_84,Com_89}, from 1999 up to 2001 with \beppo, \rxte\ and \xmm\ \citep[see e.g.][]{Zand_99, Wachter_00, Ost_01} and from 2015 up to 2017 with \swift\ \citep{Bahram_16} and \nustar\ \citep{sharma_18}.
\begin{table*} [!htbp]
\centering
\scriptsize
\begin{tabular}{ccccc}
\hline 
\hline
ObsId & Instrument & Start Time (UTC) & Exposure time (ks)\\
\hline
0008610701 & \xmm & 2001-02-20 8:28:27 & 31.5 \\
0748391601 & \xmm & 2015-09-26 19:53:05 & 42.9\\
90101013002 & \nustar & 2015-09-28 21:51:08 & 51.5\\
90201017002 & \nustar & 2016-04-21 14:41:08 & 26.8\\
00034002036\_roll1 & \swift & 2016-04-20 01:47:54 & 0.67\\
00034002036\_roll2 & \swift & 2016-04-20 01:47:54 & 0.13\\
00081918001 & \swift & 2016-04-21 20:39:01 & 0.70\\
\hline
\hline
\end{tabular}
\caption{\scriptsize Studied observations of the source \source}
\label{tab:obs}
\end{table*}

Studying the \xmm\ spectrum of \source, \cite{Sidoli_01} detected two absorption lines at 6.64 and 6.90
keV associated with the presence of highly ionised iron (\ion{Fe}{xxv} and \ion{Fe}{xxvi} ions) as well as absorption lines associated with highly ionised oxygen and neon. Furthermore, the authors detected the presence of a broad emission line centred at 6.47 keV and with a FWHM of 1.4 keV. 
Recently, \cite{Iaria_18} and \cite{Chetana_17}, studying the eclipse arrival times of the source, suggested the presence of a third body orbiting around the binary system.

Here, we report the  broadband spectral analysis of the persistent emission of \source \ using \xmm \ \citep[including the observation studied by][]{Sidoli_01}, \nustar \ and \swift/XRT data. We analysed the spectrum of the source in soft state (SS) and hard state (HS) finding that a relativistic reflection component is necessary to describe the spectra.

\section{Observations and Data Analysis}
We were interested in the spectral analysis of the persistent emission, then we extracted the \xmm/EPN, \xmm/RGS and \nustar \ events excluding the times in which the dips, the eclipses and the type-I X-ray bursts occurred. 
The analysed observations are summarised in \autoref{tab:obs}.

\begin{table*} [!htbp]
\centering
\scriptsize
\begin{tabular}{llcccc}
\hline
\hline
Model & Component & \multicolumn{2}{c}{Soft State} & \multicolumn{2}{c}{Hard State}\\
 & & \xmm\ & \swift-\nustar\ & \xmm & \nustar \\
 \hline
 {\sc Edge} &
 E (kev) & \multicolumn{2}{c}{0.538 (frozen)} & \multicolumn{2}{c}{0.538 (frozen)}\\
  & $\tau$ & \multicolumn{2}{c}{$0.24 \pm 0.03$} & \multicolumn{2}{c}{$0.30 \pm 0.08 $}\\ \\
 
 {\sc zxipcf} &
 $N_H$ (10$^22$ atoms cm$^{-2}$) & \multicolumn{2}{c}{$57^{+6}_{-13}$} & \multicolumn{2}{c}{$12.5 \pm 1.2 $} \\
& $\log \xi_{\rm IA}$  &  $4.36 \pm 0.04$& $4.13 \pm 0.07$ & $2.02 \pm 0.05$ & $3.3 \pm 0.3$\\
&  $f$ & \multicolumn{2}{c}{1 (frozen)} & \multicolumn{2}{c}{ $0.27 \pm 0.02$} \\ \\
 
 {\sc TBabs} &
 $N_H$ (10$^22$ atoms cm$^{-2}$) & \multicolumn{2}{c}{ $0.280 \pm 0.013$} & \multicolumn{2}{c}{$0.31 \pm 0.02$ }\\ \\
 
 {\sc diskbb} &
 $T_{\rm in}$ (keV) & \multicolumn{2}{c}{$0.27 \pm 0.02$} & \multicolumn{2}{c}{-} \\
& $R_{\rm in}\sqrt{\cos \theta}$ (km) & \multicolumn{2}{c}{$50^{+5}_{-10} $} & \multicolumn{2}{c}{-} \\ \\
 
 {\sc rdblur}& 
 Betor10 & \multicolumn{2}{c}{$<-2.4$} & \multicolumn{2}{c}{ $-2.5 \pm 0.2$} \\
& $R_{\rm in}$ ($R_g$) & \multicolumn{2}{c}{$39^{+35}_{-15}$} & \multicolumn{2}{c}{ $<7$ } \\ \\
 
 {\sc rfxconv} &
 rel\_refl$=\Omega/2\pi$ & \multicolumn{2}{c}{$0.30 \pm 0.08$} & \multicolumn{2}{c}{$0.48 \pm 0.06$} \\
& $\log \xi$ & \multicolumn{2}{c}{$2.72^{+0.07}_{-0.10}$} & \multicolumn{2}{c}{$1.99^{+0.04}_{-0.11}$ } \\ \\
 
 {\sc NthComp} &
 $\Gamma$ & $1.70 \pm 0.02$ & $2.153^{+0.019}_{-0.014}$ & \multicolumn{2}{c}{$2.00 \pm 0.02$}  \\
& $kT_{\rm e}$ (keV) &$2.01 \pm 0.04$ & $3.64 ^{+0.06}_{-0.08}$ & \multicolumn{2}{c}{$>150$} \\
& $kT_{\rm bb}$ (kev) & $0.42\pm 0.02$ &  $0.565^{+0.018}_{-0.007}$ & \multicolumn{2}{c}{$<0.10$}\\\\
& Norm & \multicolumn{2}{c}{$0.14 \pm 0.02$} & \multicolumn{2}{c}{$0.0341 \pm 0.0011$} \\ \\
 
& $\chi^2$(d.o.f) & \multicolumn{2}{c}{ 3124(2795)} & \multicolumn{2}{c}{2321(2269)}\\
\hline
\hline
\end{tabular}
\caption{\scriptsize Best-fit parameters for the soft state and the hard state.}
\label{tab:fits}
\end{table*} 

Analysing the RXTE/ASM and MAXI/GSC light curves of the source, we observe that the source shows a flux of 30 mCrab  during the \xmm\ obs. 0008610701, the \swift\ observations and the \nustar\ obs. 90101017002.   On the other hand, the \xmm\ obs. 0748391601 and the \nustar\ obs. 90101013002 were taken while the source flux was less than 5 mCrab.
These evidences led us to fit together the persistent spectra obtained with \xmm\ obs. 0748391601 and
\nustar\ obs. 90101013002 (hereafter HS spectrum), while the persistent spectra obtained from the \xmm\ obs. 0008610701, the \nustar\ obs. 90101017002 and the \swift\ observations are fitted together even if they were taken during two different outbursts (hereafter SS spectrum).
We will show that they describe the hard and the soft state of \source, respectively.

The adopted energy ranges for the SS are 0.4-2 keV, 0.6-12 keV, 3-35 keV and 0.5-9 keV, for RGS12, EPN, \nustar\ and \swift\ spectra, respectively; while the adopted energy range for the HS are 0.45-2 keV, 0.6-12 keV and 3-55 keV for RGS12, EPN and \nustar\ spectra.

After fitting the SS spectrum with several models, we found that the best-fit one was composed of a disc-blackbody component ({\sc diskbb} in XSPEC)  plus a thermally Comptonised component \citep[{\sc nthComp} in XSPEC, see][]{Zic_99} to fit the continuum emission, plus a convolution component which takes into account the reflection continuum also \citep[{\sc rxfconv}, see][for details]{Done_06,Kole_2011}. Furthermore, we added an absorption edge with threshold energy fixed at 0.538 keV and a  multiplicative component that takes into account a partial covering of ionised absorbing material \citep[{\sc zxipcf}, see][]{Reeves_08,Miller_07}.
We kept fixed the value of the redshift $z$ to zero, the iron abundance to the solar one, the outer radius of the accretion disc R$_{\rm out}$ to 2800 gravitational radii ($R_g=GM_{\rm NS}/c^2$) and the cosine of the inclination angle $\theta$ of \source\ \citep[that we assume to be 72$^{\circ}$,][]{Iaria_18}. The other parameters were left free to vary.

For the HS spectrum, the best-fit model was the same described above without any thermal component, which resulted unnecessary to fit the spectrum. In this case, the outer radius R$_{\rm out}$ was kept fixed to 290 $R_g$.

\begin{figure*}[!htbp]
\centering
 {  \includegraphics[scale=0.43]{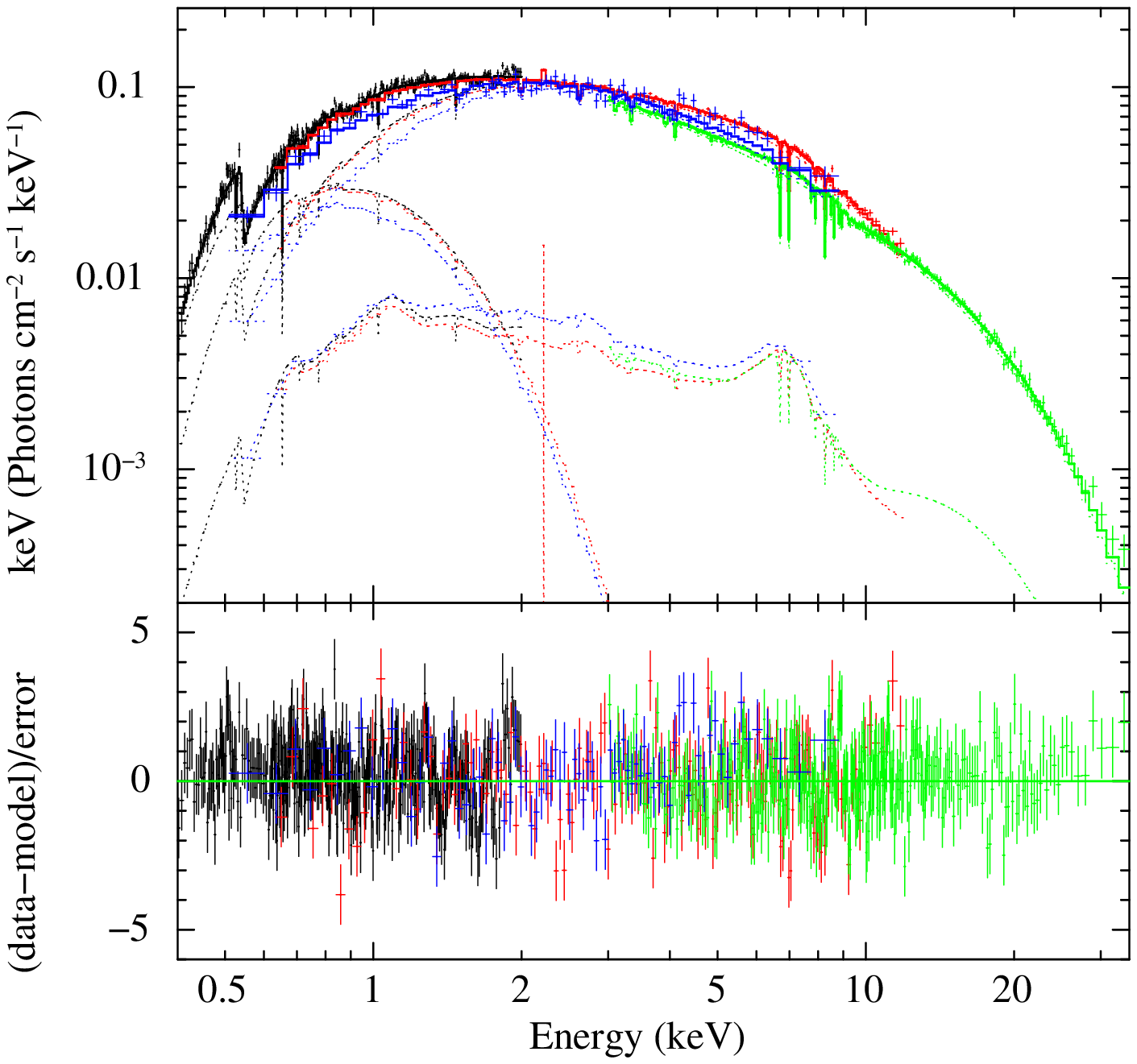}
 \hspace{0.5cm}
   \includegraphics[scale=0.43]{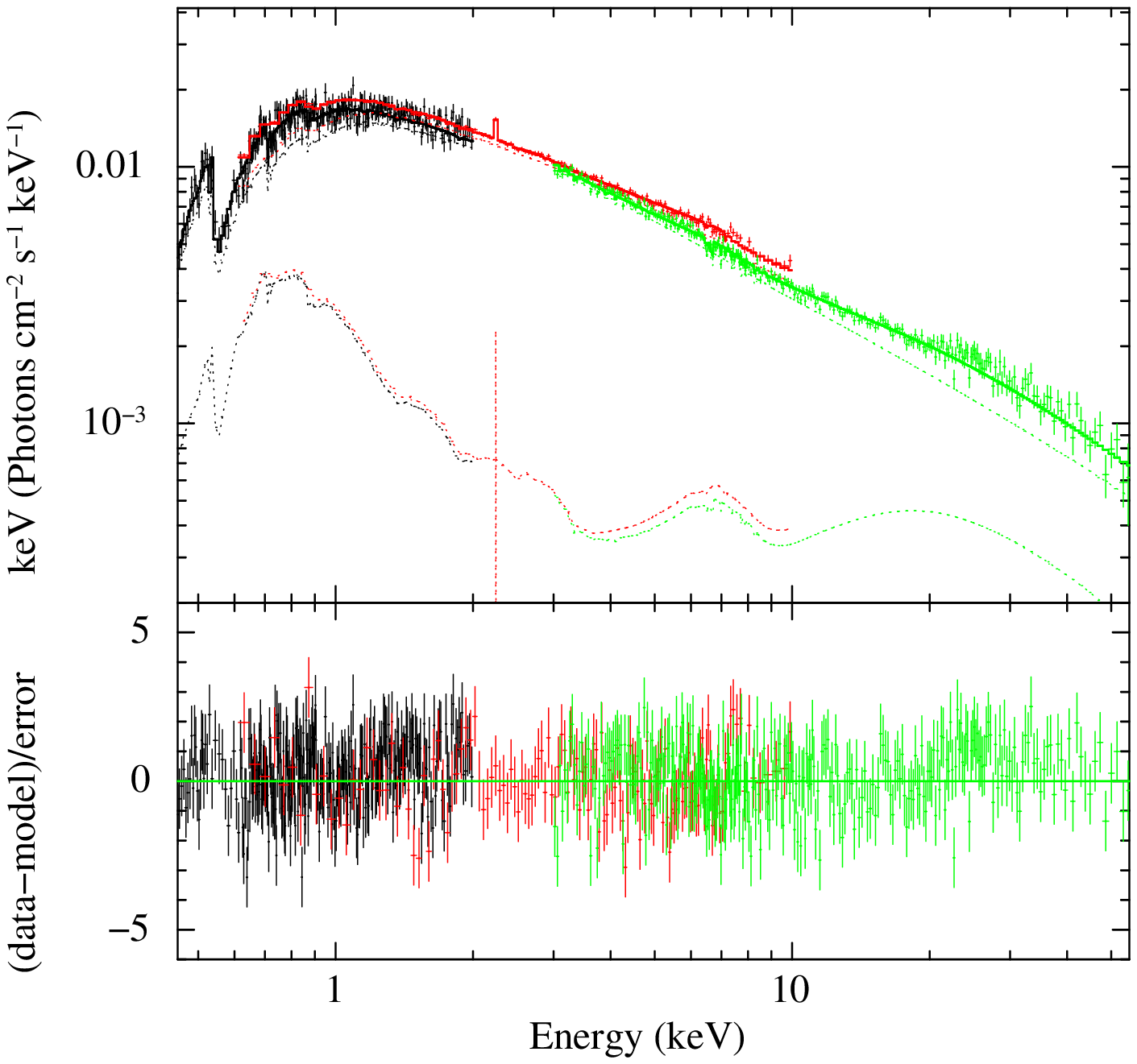}}
 \caption{\small Left Panel: SS spectrum and residuals corresponding to the best-fit model. The black, red, green and blue data are associated with the RGS12, EPN, \nustar\ and \swift\ observations, respectively.  
  Right Panel: HS spectrum and residuals corresponding to the best-fit model. The black, red and green data are associated with the RGS12, EPN and \nustar\ observations, respectively.}
              \label{fig:burst}%
    \end{figure*}

We obtained a $\chi^2(d.o.f)$ of 3124(2795) and 2321(2269), respectively; we show the best-fit values in \autoref{tab:fits}.
The unfolded spectrum and the corresponding residuals are shown in \autoref{fig:burst}.
\section{Discussion}

We have analysed several spectra of \source\ collected during the 1999 and 2015 outbursts, observing two different spectral state of the source. We find that the SS spectrum has a 0.1-100 keV unabsorbed flux of $2.2 \times 10^{-9}$ erg cm$^{-2}$ s$^{-1}$ and it shows a soft Comptonised component with a value of the electron temperature lower than 4 keV. Furthermore, a multicolour disc component is present at low energies with an inner temperature of 0.27 keV and an inner radius of the disc R$_{\rm in}\sqrt{\cos \theta} = 50^{+5}_{-10}$ km . 
On the other hand, the HS spectrum shows a hard Comptonised component with a value of the electron temperature larger than 150 keV; the addition of a thermal component is not statistically significant.
It is necessary to take into account a smeared reflection component to model the Fe-K region of both the spectra, where a broad emission line is observed. We find that the width of the emission line cannot be explained considering only the Compton scattering but we have to include the relativistic smearing in order to obtain a good fit.
In the SS, the reflecting region of the accretion disc is between 39 and 2800 $R_g$ and it is between 6 and 290 $R_g$ in HS. Besides, the ionisation parameter $\log \xi$ is $2.72^{+0.07}_{-0.10}$ and $1.99^{+0.04}_{-0.11}$ in SS and HS, respectively.
By analysing both the spectra we obtain that the values of the equivalent hydrogen column density are compatible with each other within  90\% c.l. with a value of $0.29 \times 10^{22}$ cm$^{-2}$. Furthermore, we observed that the source is covered by ionised absorbing matter, at least partially in the case of HS spectrum.

We estimated also the optical depth of the Comptonised cloud using the relation between the power-law photon index $\Gamma$ and the electron temperature $kT_{\rm e}$ obtained by \cite{Zidi_96}, finding that it is optically thick in the SS (i.e $8 < \tau < 16$) whilst it is optically thin in the hard state ($\tau < 0.7$).
Finally, we obtained a value of the relative reflection normalisation of $0.30 \pm 0.08$ in SS, compatible with a spherical corona in the inner part of the accretion disc \citep[see][]{Dove_97}; while the value of $ 0.48 \pm 0.06$ obtained for the HS suggests a larger superposition of the corona to the disc.

\section*{Acknowledgements}
We acknowledge financial contribution from the agreement ASI-INAF I/037/12/0.

\bibliographystyle{aa}
\bibliography{biblio}
\end{document}